# THE RETURN TO HIGHER EDUCATION: EVIDENCE FROM ROMANIA

Bogdan OANCEA, Richard POSPÍŠIL, Raluca Mariana DRĂGOESCU

University of Bucharest - Faculty of Administration and Business - Department of Economic and Administrative Sciences, Palacky Univeristy of Olomouc - Faculty of Arts - Department of Applied Economics, National Statistics Institute of Romania

**Abstract**: Education is one of the most important components of the human capital, and an important determinant of the personal income. Estimating the rate of return to education is a main topic of economic research. In this paper we analyzed the rate of return to higher education in Romania using the well-known Mincer equation. Besides the educational level and the number of years of experience on the labor market we also used a series of socio-demographic variables such as gender, civil status, the area of residence. We were interested mainly in calculating the rate of return to higher education, therefore we computed this rate for bachelor, master and doctoral degrees separately. We also investigated the rate of return to higher education on technical, science, economics, law, medicine, and arts fields. Our results showed that the rate of return to higher education has a greater value than most of the developed countries of EU and the field of higher education that brings the highest rate of return is medicine.

## 1 Introduction

The labor market economy has undergone a major transformation during the last century. If in the beginning, labor as a factor was regarded as a homogeneous conglomerate of workers who became input into a production function across the economy, modern labor market approaches see work as a heterogeneous conglomerate of people with different labor productivity. As a result, economists studying the labor market are now trying to find a distribution of workers' income rather than studying how the economic outcome between labor and capital is distributed.

Mincer (1958) was one of the first economists who used the term human capital in his pioneering studies in which he attempted to model income distribution using mathematical tools of the neoclassical theory of capital. Using a simple model that involved only the number of years of education and then the age and number of weeks worked per year by a person, Mincer was able to explain about 60% of the annual income variation for the US white population. This model was then applied in over 100 countries with the same success. According to this model, the rate of return to education was found to take values between 5% and 15%, just like commercial investments. He introduced the observation that an individual's choices in terms of labor options produce an income flow that can easily be measured using capital theory. If education and occupation on the labor market are treated as investment opportunities, the outcomes of investments made by an individual can be easily modeled. Assuming that people invest in education to the point where the cost of investment equals the present value of education gains, Mincer has achieved an econometric model that shows that a person's income increases at a decreasing rate over the lifetime, leading to a concave function of time.

Mincer (1958, 1974, 1978) managed to derive an empirical formula for a person's income over his lifetime. At any time *t* of a person's life, incomes can be viewed as a linear function of schooling years and concave of the number of years of experience on the labor market. This equation, called Mincer equation, is shown below:

$$\ln(Y_i(S,t)) = b_0 + b_1 S_i + b_2 t_i + b_3 t_i^2 + \varepsilon_i \qquad (1)$$

where $Y_i(t)$ is the income of the person *i*, $S_i$ is the number of years of education and $t_i$ is the number of years spent on the labor market. The coefficient $b_0$ represents the initial capacity of earning money, $b_1$ the rate of return to education, and coefficients $b_2$ and $b_3$ are related to the amount and rate of financial return on workplace training.

There are several empirical implications of the Mincer equation. The first one is that the income is directly correlated with the investment in human capital. The coefficient of the schooling variable in the Mincer equation is the internal rate of return to education. This can be understood as the discount rate at which education costs during the study period seen as opportunity costs are matched by future earnings. The second observation relates to the fact that the income function is concave on the number of years of experience on the labor market, which means that the income grows more quickly for young people, reaches a maximum, then it starts to decrease.

The rate of return to education has been calculated for many countries and time periods. Psacharopoulos and Patrinos (2004) presented the results obtained for 70 countries for a period of 25 years and they found that the rate of return on education has values between 5% and 17%. Weisberg (1995) studied the evolution of the rate of return to education in Israel using data covering the period 1974-1983 and noted that it had an increasing trend over time. It also showed that the rate is higher for higher education graduates. Another study (Trostel, Walker and Woolley, 2002) presented estimates of the rate of return to education for 28 countries between 1980 and 2000 using different data sources, obtaining values between 3% and 17%. A conclusion that comes from these studies is that the rate of return to education for most of the developed countries has lower values than for developing countries. While this rate ranges from 5% to 10% for most of the developed countries, it has greater values (10% - 17%) for the developing countries.

The rest of the paper is organized as follows. In the next section we present some methodological problems regarding the estimation of Mincer equation, in section 3 the estimates for Romania, in section 4 a discussion on the obtained results and section 5 concludes our paper.

## 2 Methodological issues regarding Mincer equation

In this section we present the main difficulties when estimating Mincer equation with empirical data:

a) Non probabilistic samples. When the estimation is performed using a non-probabilistic sample the estimators are usually biased.

b) Omission of some variables. If some variables are omitted from the equation, the results will be biased when these variables are correlated with the dependent variable and other independent variables.

c) The functional form of the equation. Equation (1) has a log-linear form and experimental studies performed so far show that it is the best fit for experimental data, but cases have also been reported where the log-log function was better (Thurow, 1969).

d) Unobserved heterogeneity. In the case of multivariate regression, it is often the case when important variables are omitted because there are no data for such variables. For example, one of the important variables in determining a person's income is his/her innate ability. In this case, the rate of return to education will be overestimated. A solution to this problem is the use of instrumental variables and the two-stage least squares method (2SLS).

e) Endogeneity of the schooling variable. The issue of endogeneity occurs when certain determinants of the response variable are also correlated with the exogenous variables and they are not directly observable. As a rule, a person's education can be influenced by a number of factors such as the level of education of his/her family, living conditions, etc. It is obvious that different people cannot be identical regarding some unobservable variables. In such situations where a variable influences both the income and education, the estimation of Mincer equation through OLS method results in biased estimates (Card, 2001). This bias can be eliminated using instrumental variables and 2SLS estimation method. Below are some instrumental variables found in the literature:

- Variables related to the family background (Trostel et al., 2002; Zhang, 2011; Liu et al., 2000);

- Proximity of educational institutions (Card, 1993; Warunsiri, 2010; Flabbi,1999);

- Other instrumental variables: the minimum legal age when leaving the school is allowed (Harmon and Walker, 1995), the reforms in the educational system (Brunello and Miniaci,1999; Ismail, 2007).

Although there have been many proposals for instrumental variables in the studies so far, finding suitable instrumental variables is still an area of research. Instrumental variables that are even very poorly correlated with the endogenous variable lead to biased results even for very large samples (Bound, 1995).

## 3 Estimation of the Mincer equation for Romania

We used the 2011 Population and Housing Census data and personal gross income tax records for 2013 to estimate Mincer equation for Romania. To our knowledge this is the first study at this level of detail in Romania and one of the fewest studies for Romania.

Ion (2013), using the data from the 2009 Household Budget Survey, estimated the impact of education on income and concluded that the rate of return to education is 11.29% which is comparable to that obtained by Mincer (1974) 10.7% and according to (Psacharopoulos and Patrinos, 2004) is higher than the value for several developed countries, which means that in Romania the labor market rewards education more than other countries.  Ion used a representative sample at national level, and estimated a regression equation for income based on a series of explanatory variables by OLS. The following explanatory variables were used: the number of years of education, the labor market experience, gender, nationality, civil status, the sector where the person works (public or private), the field of activity, development region, area of residence (urban / rural), squared number of years of experience.

A second study on the Mincer equation for Romania is Pauna (2009) who estimated the equation for 1995 and 2000 using OLS and showed that the rate of return to education had a negative sign that contradicts Mincer's theory. This coefficient calculated separately for men and women has positive values (5.9%, respectively 7.18% in 1995 and 8.3% respectively 9.75% in 2000) with an increasing trend, which means that in Romania the labor market has increased the reward for education from 1995 to 2000.

The third study (Varly et al., 2014) used data available from the SILC and Household Bugdets surveys from 2012 to estimate the Mincer equation. The results showed that the rate of return to education is 7.9% when using the SILC data, and 8.6% for the HB survey. Varly (2014) also estimated the equation by the level of education (primary, secondary, bachelor or doctorate) and obtained a wage premium of approximately 51% for bachelor and 42% for doctoral graduates.

Our analysis is based on data from the 2011 Population and Housing Census. Unlike other studies using only a sample, we used an exhaustive dataset covering the entire population of Romania.

Income, education and labor market experience are the fundamental variables of the Mincer model. For education we used the number of schooling years computed from the census data using the conventions presented in Tables 1 and 2. For incomes, we used the information from the tax records (2013) containing the gross annual incomes.

**TAB. 1: Number of years of education for higher education graduates**

| Level of Education | Number of years of education |
|---|---|
| Bachelor | 16 |
| Masters | 18 |
| Doctoral education | 21 |

**TAB. 2: Number of years of education for secondary and primary education levels**

| Level of Education | Number of years of education |
|---|---|
| Secondary education (high school) | 12 |
| Postsecondary education and foremen schools | 15 |
| Vocational education | 10 |
| Gymnasium education | 8 |
| Primary education | 4 |

It should be noted that the number of years of education used here may differ from theirs real value especially in higher education because there are persons who have repeated years of study or who have completed their studies faster. Also, different areas of study require different number of years. Labor market experience is an essential variable in the Mincerian model, but it is not included in the data collected at the Census. Many empirical studies use age as a proxy for experience (Sanroman, 2006; Hyder, 2007), but this can lead to biased estimates especially for young people. We used the potential labor market experience calculated according to the following formula:

$$EXP = AGE - EDU - 6 \tag{2}$$

In addition to these fundamental variables of the Mincer model, we also used the following control variables:

GENDER: 1 for men, for women;

MARRIED: 1 if the person is married, 0 otherwise;

WTIME: time worked - the number of hours worked per week multiplied by the number of weeks the person earned money in the year under consideration;

BIG_TOWN: 1 if the person lives in a city with more than 150,000 inhabitants, 0 otherwise;

URBAN: 1 if the person resides in the urban area, 0 otherwise.

To characterize the rate of return to higher education, which is one of the main purposes of this paper, we used the following dummy variables:

HAS_HS: 1 if the person has high school education, 0 otherwise;

HAS_HE: 1 if the person has a bachelor degree, 0 otherwise;

HAS_MA: 1 if the person has a master degree, 0 otherwise;

HAS_DR: 1 if the person has a doctorate, 0 otherwise;

HE_TECH: 1 if the person has technical higher education, 0 otherwise;

HE_UNIV: 1 if the person has science higher education, 0 otherwise;

HE_EC: 1 if the person has economics higher education, 0 otherwise;

HE_LAW: 1 if the person has law higher education, 0 otherwise;

HE_MED: 1 if the person has medical higher education, 0 otherwise;

HE_ART: 1 if the person has higher education in the arts, 0 otherwise;

We included all persons aged between 15 and 64, employed on the labor market who earned an income in the reference year. We started by estimating the following equation:

$$LOG\_V_i = b_0 + b_1 EDU_i + b_2 EXP_i + b_3 EXP_i^2 + b_4 GENDER_i + b_5 MARRIED_i + b_6 WTIME_i + b_7 BIG\_TOWN_i + \varepsilon_i \qquad (3)$$

The results are presented in table 3.

**TAB. 3: The results of estimating the Mincer equation by OLS method**

| Variable | Coefficient |
| --- | --- |
| Education | 0.1228 |
| Experience on labor market | 0.0197 |
| Squared number of years of experience on labor market | -0.0002 |
| Gender | 0.0912 |
| Marital status | 0.0100 |
| Number of weekly hours worked | 0.0010 |
| Living in a big city | 0.1487 |

* $R^2$=0.42, adj $R^2$=0.42, coefficients are significant at 0.0001

The rate of return to education is 12.28%, i.e. an additional year of education brings an increase of the income by 12.28%. Compared with previous results obtained in the case of Romania, there is a substantial increase. One year of additional experience adds 1.9% to the income. The coefficient of EXP2 is negative which is in line with Mincer's theory.

The coefficient of the GENDER variable shows that men earn 9.1% more than women and living in a city with more than 150,000 inhabitants leads to an income with 14.8% greater than the rest of the employees. Married people earn 1% more while the WTIME variable shows that one extra hour of work per year leads to a 0.1% increase of the income.

The value $R^2$ = 0.42 is comparable to that obtained by Mincer in his initial study (Mincer, 1974) and higher than those obtained by Ion (2013) or Pauna (2009).

The bias of estimators due to the endogenousness of the education may result either from the unobservable variation of employee' ability or from unobserved heterogeneity. For example, there may be people who continue their education after completing the formal education period, thus gaining superior abilities. The estimator of the rate of return to education becomes biased, the labor productivity being different in this case due to the skills acquired outside the formal education period. To address this issue we used instrumental variables and estimation using the 2SLS method.

We used the environment in which the person resides as an instrument (the URBAN variable). Estimating the coefficients of the Mincer equation by 2SLS method consists in the following steps:

1. The coefficients of the equation with education as an endogenous variable are estimated with OLS:

$$\text{EDU}_i = c_0 + c_1 EXP_i + c_2 EXP_i^2 + c_3 GENDER_i + c_4 MARRIED_i + c_5 WTIME_i + c_6 BIG\_TOWN_i + c_7 URBAN_i + \varepsilon_i \qquad (4)$$

2. The coefficients of the initial Mincer equation in which the variable $\text{EDU}_i$ is replaced by the fitted values given by the equation (4) $\widehat{EDU}_i$ are then estimated:

$$\text{LOG\_V}_i = b_0 + b_1 \widehat{EDU}_i + b_2 EXP_i + b_3 EXP_i^2 + b_4 GENDER_i + b_5 MARRIED_i + b_6 WTIME_i + b_7 BIG\_TOWN_i + \varepsilon_i \qquad (5)$$

The results are presented in table 4.

**TAB. 4: The results of estimating the Mincer equation using 2SLS method**

| Variable | Coefficient |
| --- | --- |
| Education | 0.1610 |
| Experience on labor market | 0.0258 |
| Squared number of years of experience on labor market | -0.0002 |
| Gender | 0.0912 |
| Marital status | -0.0128 |
| Number of weekly hours worked | 0.0010 |
| Living in a big city | 0.1014 |

* $R^2$=0.42, adj $R^2$=0.42, coefficients are significant at 0.0001

These results confirm that the rate of return to education is underestimated by the OLS method compared to 2SLS: an extra year of education leads to a 16.1% increase to the income, and an additional year of experience adds 2.5%, higher than the OLS estimate. The coefficient of the GENDER variable shows that men earn 11.7% more than women. Gender differences are also present in studies for other countries (Fiaschi and Gabbriellini, 2013).

Living in a city with more than 150,000 inhabitants leads to an income 10.1% greater than the rest of the employees. Married people earn 1.2% less, while the WTIME variable shows that one extra hour of work per year leads to a 0.1% increase of the income.

Testing that the EDU variable is indeed endogenous was achieved using the Hausman test.

**TAB. 5: The Hausman test for detecting endogeneity of the EDU variable using the URBAN variable as an instrument**

| Efficient under H0 | Consistent under H1 | Hausman statistics | p-value |
|---|---|---|---|
| OLS | 2SLS | 2580.88 | 0.0000 |

The very low p-value suggests that the null hypothesis that there are no significant differences between OLS and 2SLS estimates is rejected. This means that EDU variable is not exogenous, leading us to conclusion that 2SLS method for estimating the coefficients of the Mincer equation is more robust compared to the OLS.

A problem that often occurs when using instrumental variables is the use of weak instruments, i.e. variables that are poorly correlated with the endogenous variable, which will ultimately lead to biased estimators in the same direction as those obtained by OLS (Bound et al., 1995). Staiger and Stock (1997) set out an empirical rule to test whether an instrument is weak or not: a partial F-test in the first stage of the 2SLS method calculated by excluding the instrument less than 10 indicates a weak instrument. We applied this test and obtained a value much greater than 10 indicating that the instrument we used is strongly correlated with the education variable.

We will relax the hypothesis that the rate of return to education is constant and allow this rate to vary according to the educational level by replacing the schooling variable with the dummy variables HAS_PROF, HAS_HS, HAS_POST, HAS_HE, HAS_MA, HAS_DR. The use of dummy variables to highlight the level of education is in line with modern human capital theories that states that a person's income does not depend so much on how many years he/she spent in school, but on the earned diploma. This hypothesis is based on the observation that in the presence of heterogeneity, what matters is the type of graduating institution and not the number of years of study. We will estimate the following model:

$$LOG\_V_i = b_0 + b_1 HAS\_PROF_i + b_2 HAS\_HS_i + b_3 HAS\_POST_i + b_4 HAS\_HE_i + b_5 HAS\_MA_i + b_6 HAS\_DR_i + b_7 EXP_i + b_8 EXP_i^2 + b_9 GENDER_i + b_{10} MARRIED_i + b_{11} WTIME_i + b_{12} BIG\_TOWN_i + \varepsilon_i \quad (6)$$

**TAB. 6: The results of estimating Mincer equation using different levels of education**

| Variable | Coefficient |
|---|---|
| HAS_PROF | 0.129 |
| HAS_HS | 0.274 |
| HAS_POST | 0.598 |
| HAS_HE | 0.944 |
| HAS_MA | 1.167 |

| HAS_DR | 0.977 |
|---|---|
| EXP | 0.023 |
| EXP2 | -0.00029 |
| MARRIED | 0.012 |
| GENDER | 0.099 |
| BIG_TOWN | 0.144 |

\* $R^2$=0.43, adj $R^2$=0.43, coefficients are significant at 0.0001

This result indicates that the labor market in Romania rewards higher education: higher education employees earn much higher incomes than the rest of the employees. Similar results are obtained by Fiaschi and Gabbrilellini (2013) for Italy or Humphreys (2013) for Cambodia in 2012.

If we convert these coefficients of the dummy variables into a relative effect on income, following the methodology described in (Kifle, 2007) and (Halvorsen and Palmquist, 1980), we obtain the results presented in table 7.

**TAB. 7: Relative effect of education level on income (reference level - gymnasium studies)**

| Level of education | The relative effect of the level of education on income |
|---|---|
| Vocational education | 13.7% |
| High school | 31.6% |
| Post-secondary education | 81.9% |
| Bachelor degree | 157.2% |
| Masters degree | 221.5% |
| Ph.D. degree | 165.5% |

The graduates of vocational education have a gain of 13.7% compared to those with gymnasium, the high school graduates 31.3%, the post-secondary education 81.9%, the undergraduate higher education 157.2% , masters higher education graduates have a gain of 221.5% and those who have doctorates earn 165.5% more. These results show that the rate of return to education is not constant across all levels of education as in Mincer's initial model, but it varies according to the educational level, the highest value being met for higher education. The graduates of a doctoral program register a slight decrease compared with bachelor and masters graduates explained probably by the fact that most people with a Ph.D. degree work either in the educational or research system where the incomes are low. These results are in line with other recent studies (Belzil, 2006), (Varly et al., 2014), (Psacharopoulos, 2004).

Transforming these rates into annualized rates is straightforward (Kifle, 2007) and the annualized rates of return to education are presented in table 8.

**TAB. 8: Annualized rate of return on education by educational level**

| Level of education | Annualized rate of return to education |
|---|---|
| Vocational education | 1.37% |
| High school | 2.36% |
| Post-secondary education | 5.42% |
| Bachelor degree | 9.83% |
| Masters degree | 12.30% |
| Doctoral degree | 7.88% |
| High school vs vocational education | 7,8% |
| Post-secondary education vs high school | 16.9% |
| Bachelor degree vs post-secondary education | 41.4% |
| Masters degree vs bachelor degree | 12.5% |
| Doctoral degree vs masters degree | -5.8% |

One can notice that the annualized rate of return to education increases with the increase of the educational level, the highest value being recorded for graduates of higher education compared to those with post-secondary education, 41.4%. The increase of the rate of return to higher education can be noticed as well as for the master degree compared to the bachelor, but this is only 12.5%. The doctorate degree does not bring anything in addition to the masters, on the contrary the annualized rate is negative -5.8%.

We will refine the study on the influence of graduating a form of higher education on income by further detailing the field of higher education: technical, science, economics, law, medicine and arts. We estimated the equation:

$$LOG\_V_i = b_0 + b_1 HE\_TECH_i + b_2 HE\_UNIV_i + b_3 HE\_EC_i + b_4 HE\_LAW_i + b_5 HE\_MED_i + b_6 HE\_ART_i + b_7 EXP_i + b_8 EXP_i^2 + b_9 GENDER_i + b_{10} MARRIED_i + b_{11} WTIME_i + b_{12} BIG\_TOWN_i + \varepsilon_i \quad (7)$$

by OLS and the results are presented in table 9. The field that brings the greatest benefit on the labor market is medicine. Graduates of technical and economics education follow, and then graduates of law studies. For economics education the results are somewhat normal given the evolution of the higher education and labor market in Romania after 1990 (Andrei et al. 2010a; Andrei et al., 2010b; Dragoescu, 2015), but for the technical education, the results may be surprising given the deindustrialization of Romania after 1990. It is possible that the explanation comes from the fact that the income in IT and telecommunications where large numbers of graduates of technical education work are very high.

**TAB. 9: Estimation of Mincer equation by the field of higher education using OLS**

| Variable | Coefficient |
|---|---|
| HE_TECH | 0.778 |
| HE_UNIV | 0.719 |
| HE_EC | 0.776 |
| HE_LAW | 0.725 |
| HE_MED | 0.853 |
| HE_ART | 0.601 |
| EXP | 0.023 |
| EXP2 | -0.0003 |
| GENDER | 0.079 |
| MARRIED | 0.026 |
| WTIME | 0.001 |
| BIG_TOWN | 0.164 |

The least "profitable" fields of higher education are science and arts. This is normal, since the graduates from science faculties have as their main employer the educational and research system characterized by low incomes and the graduates from arts also have low incomes.

The coefficients of the dummy variables are converted to annual rates and presented in Table 10. We considered two variants to calculate these rates: the same duration of studies in all fields and different study duration.

In the first variant, the highest rate is recorded for medicine graduates (33.67%) followed by the technical and economics (about 29%), then the science education with 26.33% and finally the arts. If we consider that the time required to graduate in different fields of higher education differs, the rate of return to education is highest in economics followed by technical education. Medical studies have a lower rate of return, explained by the fact that these studies last for 6 years.

**TAB. 10: Annualized rate of return on higher education by the field of study**

| Field of study | Annualized rate of return to education (the same duration of studies) | Annualized rate of return to education (different duration of studies) |
|---|---|---|
| Technical | 29.48% | 29.48% |
| Economics | 29.34% | 39.12% |
| Medicine | 33.67% | 22.45% |
| Law | 26.65% | 26.65% |
| Science | 26.33% | 26.33% |
| Arts | 20.59% | 20.59% |

## 4 Conclusions

In this paper we studied the influence of education on the income in Romania using the well-known Mincer equation. While this subject attracted many studies at international level, in Romania there are only few papers, our study being the single one that used an exhaustive database for the whole population of the country. We showed that the rate of return to education has increased over the time compared with some previous studies. We also showed that using OLS the estimates are downward biased: while the rate is 12.28% computed by OLS, it is 16.1% when we used an instrument and 2SLS. A special attention was paid to the higher education showing that it brings greater benefits on the labor market compared with other educational levels. We analyzed separately the bachelor, masters and doctorate degrees and showed that while bachelor and masters bring an increase to the log of income, the doctorate degree has a lower rate of return than the master degree. Analyzing the influence of the field of higher education, we showed that the areas that has the highest rate of return are medicine, economics and law.

**Literature**:

Andrei, T., Teodorescu, D., Oancea, B., Iacob, A. (2010a). Evolution of higher education in Romania during the transition period. *Procedia - Social and Behavioral Sciences*, 9, 963-967.

Andrei, T., Teodorescu, D., Oancea, B. (2010b). Characteristics of higher education in Romania during transition. *Procedia - Social and Behavioral Sciences*, 2(2), 3417-3421.

Belzil, C. (2006). *Testing the Specification of the Mincer Wage Equation*, Working Paper 06-08, CATE UMR 5824 du CNRS, Ecully. Retrieved June 3, 2016 from http://hal.archives-ouvertes.fr/docs/00/14/25/42/PDF/0608.pdf.

Bound, J., Jaeger, D. A., Baker, R. M. (1995). Problems with Instrumental Variables Estimation when the Correlation Between the Instruments and the Endogenous Explanatory Variable is Weak. *Journal of the American Statistical Association*, 90(430), 443-450.

Brunello, G., Miniaci, R. (1999). The economic returns to schooling for Italian men. An evaluation based on instrumental variables, *Labour Economics*, 6(4), 509-519.

Card, D. (1993). *Using Geographic Variation in College Proximity to Estimate the Return to Schooling*. National Bureau of Economic Research (NBER). Working Paper No. 4833.

Card, D. (2001). Estimating the Return to Schooling: Progress on Some Persistent Econometric Problems. *Econometrica*, 69 (5), 1127-1160.

Dragoescu, R.M., (2015). *Quantitative analysis of the evolution and trends of the higher education in Romania.* Ph.D. Thesis (in Romanian), The Bucharest University of Economic Studies.

Fiaschi, D., Gabbriellini, C. (2013). *Wage Functions and Rates of Return to Education in Italy.* Presented at the Fifth meeting of the Society for the Study of Economic Inequality, Bari. Retrieved on February 2, 2015 from http://www.ecineq.org/ecineq_bari13/FILESxBari13/CR2/p253.pdf.

Flabbi, L. (1999). *Returns to Schooling in Italy: OLS, IV and Gender Differences*. Bocconi University, Econometrics and applied Economics Series. Working Paper No. 1.

Halvorsen, R., Palmquist, R. (1980). The interpretation of dummy variables in semilogarithmic equations, *The American Economic Review*, 70(3), 474-475.

Harmon, C., Walker, I. (1995). Estimates of the Return to Schooling for the United Kingdom, *The American Economic Review*, 85(5), 1278-1286.

Hyder, A. (2007). *Wage Differentials, Rate of Return to Education, and Occupational Wage Share in the Labour Market of Pakistan*. Pakistan Institute of Development Economics (PIDE) Working Paper No. 2007:17.

Humphreys, J. (2013). *An alternative to the Mincer model of education*, In: 42nd Australian Conference of Economists, Murdoch University Western Australia.

Ion, I., (2003). Education In Romania - How Much Is It Worth?. *Romanian Journal of Economic Forecasting*, 1, 149-163.

Ismail, R. (2007). *The Impact of Schooling Reform on Returns to Education in Malaysia*. Munich Personal RePEc Archive (MPRA). University Library of Munich. Paper No. 15021.

Kifle, T. (2007). The Private Rate of Return to Schooling: Evidence from Eritrea. *Essays in Education*, 21, 77-99.

Liu, J.-T., Hammitt, J. K., Lin, C. J. (2000). Family background and returns to schooling in Taiwan. *Economics of Education Review*, 19(1), 113-125.

Mincer, J. (1958). Investment In Human Capital and the Personal Income Distribution. *Journal of Political Economy*, 66, 281-302.

Mincer, J. (1974). *Schooling, Experience, and Earnings.* Columbia University Press for the National Bureau of Economic Research, New York.

Mincer, J. (1978). Family Migration Decisions. *Journal of Political Economy,* 8(5), 749-773.

Păuna, C., (2009). *Wage discrimination in Romania - Evolution and explanations* (in Romanian). Working Papers of Macroeconomic Modelling Seminar, 19, 081902. Retrieved on April, 4, 2015 from http://www.ipe.ro/RePEc/WorkingPapers/cs19_2.pdf.

Psacharopoulos, G., Patrinos, H. A. (2004). Returns to Investment in Education: A Further Update. *Education Economics*, 12(2), 111-134.

Sanroman, G. (2006). *Returns to schooling in Uruguay*. Economics Department. Social Sciences Faculty, University of the Republic, Montevideo, Uruguay. Working Paper No. 14/06.

Staiger, D., Stock, J. H. (1997). Instrumental Variables Regression with Weak Instruments, *Econometrica*, 65, 557-586.

Varly, P., Iosifescu, C. S., Fartușnic, C., Andrei, T., Herțeliu, C. (2014). *Cost of non-investment in education in Romania*. UNICEF, Bucharest.

Thurow, L. (1969). *Poverty And Discrimination*. Washington, Brookings Institution.

Trostel P., Walker, I., Woolley, P. (2002). Estimates of the Economic Return to Schooling for 28 Countries. *Labour Economics*, 9(1), 1-16.

Zhang, X. (2011). The Rate of Returns to Schooling: A case study of Urban China. *International Journal of Humanities and Social Science*, 1(18), 173-180.

Warunsir, S., Mcnown, R. (2010). The Returns to Education in Thailand: A Pseudo-Panel Approach. *World Development*, 38(11), 1616-1625.

Weisberg, J. (1995). Returns to Education in Israel: 1974 and 1983. *Economics of Education Review*, 14(2), 145-154.